\documentclass[aps,prl,showpacs,amssymb,twocolumn,nofootinbib]{revtex4}
\usepackage{graphicx}
\begin{document}
\title{Time uncertainty in
quantum gravitational systems}

\author{J. Fernando \surname{Barbero G.}}
\affiliation{IMAFF, CSIC, Serrano 113bis-121, 28006 Madrid, Spain}
\author{Guillermo A. \surname{Mena Marug\'an}}
\affiliation{IMAFF, CSIC, Serrano 113bis-121, 28006 Madrid, Spain}
\author{Eduardo J. \surname{S. Villase\~nor}}
\affiliation{IMAFF, CSIC, Serrano 113bis-121, 28006 Madrid, Spain}

%\date{1 August 2003}

\begin{abstract}
It is generally argued that the combined effect of the Heisenberg
principle and general relativity leads to a minimum time
uncertainty. Most of the analyses supporting this conclusion are
based on a perturbative approach to quantization. We consider a
simple family of gravitational models, including the
Einstein-Rosen waves, in which the (nonlinearized) inclusion of
gravity changes the normalization of time translations by a
monotonic energy-dependent factor. In these circumstances, it is
shown that a maximum time resolution emerges nonperturbatively
only if the total energy is bounded. Perturbatively, however,
there always exists a minimum uncertainty in the physical time.
\end{abstract}

\pacs{04.60.Ds, 04.62.+v, 03.65.Ta, 06.30.Ft}

\maketitle

Given a quantum state, one can track the passage of time by
analyzing the evolution of probability distributions of
observables \cite{GP}. However, for every observable, there is a
characteristic time that places a limit on the ability to detect
the evolution. This characteristic time can be estimated as the
ratio between the root-mean-square (rms) deviation of the
observable and the (absolute value of the) time derivative of its
expectation value. In conservative systems, the noncommutativity
of quantum mechanics implies that, for all explicitly
time-independent observables, this characteristic time is greater
or equal than $\hbar/2$ divided by the rms deviation $\Delta H$ of
the energy \cite{GP}. As a consequence, any measurement of time
has an intrinsic uncertainty $\Delta t$ that satisfies the
so-called fourth Heisenberg relation, $\Delta t \Delta H\geq
\hbar/2$. For eigenstates of the Hamiltonian, the probability
distributions are stationary. To increase the time sensitivity,
one must allow for states with a larger and larger energy
uncertainty. A perfect time resolution can be reached only when
the energy is completely delocalized.

The discussion gets much more involved when general relativity
enters the scene. It is commonly accepted that the above quantum
mechanical description should be valid in the low-energy regime,
or around a background that provides the fundamental state.
However, higher-order corrections in this approximation should
become important when one considers states with large energy
fluctuations, necessary for a good time resolution. Indeed,
several arguments indicate that a minimum time structure appears
when one includes at least the next-to-leading order contribution
to the time uncertainty \cite{Luis,Pad}.

A way to understand this phenomenon is by the back reaction caused
by the energy of the quantum system. In general relativity, this
energy curves the spacetime. If the physical time is defined in
terms of a unit (asymptotic) timelike Killing vector, the presence
of additional energy around the background modifies the
normalization of this vector and hence the definition of time.
Since this modification depends on the amount of extra energy,
quantum uncertainties in the energy give rise to time
uncertainties \cite{Luis}. This mechanism prevents one from
attaining the limit of infinite time resolution by increasing the
energy fluctuations unless the contributions to the time
uncertainty arising from quantum mechanics and general relativity
are correlated in a very specific manner.

There exists a certain similarity between these arguments and
those supporting the existence of a minimum time (or length) in
the string \cite{Luis, ACV, GKP}. The spacetime uncertainty
resulting from scattering processes in string perturbation theory
\cite{ACV} can be understood as produced by the independent
fluctuations of the two worldsheet directions \cite{GKP}. These
fluctuations are given by two dual extremal lengths, whose product
never vanishes. In fact, the uncertainty in the time direction of
the worldsheet is proportional to that in the time-of-flight
measurement of the momentum, which increases when one improves the
resolution in the spatial direction \cite{GKP}.

This line of reasoning has led to different proposals for the
minimum time uncertainty that one should expect in gravitational
systems \cite{Pad,AC,NgVD}. The simplest proposal is an
uncertainty of Planck order \cite{Luis,Pad}. Assuming random
fluctuations at the Planck scale, an uncertainty that increases
with the square root of time has also been suggested \cite{AC}.
The same behavior was found by Salecker and Wigner (SW) by
analyzing a device acting as a clock, with initial position and
momentum rms deviations $\Delta x$ and $\Delta p$ and mass $m$
\cite{SW}. Again, as time passes, the position uncertainty
receives an energy correction \cite{SW,SWCr}, namely $[\Delta
x(t)]^2=[\Delta x]^2+[t\Delta p/m]^2$. The Heisenberg principle
implies then that the minimum of the time uncertainty $\Delta
t\equiv \Delta x(t)$ (with $c=1$) is proportional to $\sqrt{t}$
\cite{AC,SW,AC2}.

Although these proposals are not free of controversy
\cite{SWCr,SWM}, they have originated an increasing interest for
the consequences that a minimum time uncertainty of quantum
gravitational nature could have in astrophysics
\cite{NgVD,ACW,LH}. For instance, it has been proposed that this
uncertainty might cause a distinctive displacement noise in
gravitational wave interferometers \cite{AC,ACW}. Another effect
would be the loss of phase coherence in the radiation emitted by
distant astrophysical sources, which would prevent the formation
of diffraction patterns \cite{LH,RTG}. This last suggestion has
received several criticisms \cite{AsCr}.

These predictions are sometimes regarded as observational tests of
the time uncertainty in quantum gravity. However, they are deduced
in fact under stronger hypotheses, which imply a foamy spacetime.
Thus, for gravitational wave detectors a fuzzy concept of distance
is assumed \cite{AC}. On the other hand, the light coming from
extragalactic objects suffers a significant loss of phase only if
there exist spacetime fluctuations which induce random phase
variations along the propagation path \cite{RTG}.

The extent to which a minimum time uncertainty leads to testable
consequences is obscured by the use of descriptions that are
mainly phenomenological, rather than obtained from a consistent
quantization in the presence of gravity. Moreover, most of the
results that support the existence of a minimum time structure are
based on qualitative analyses that involve perturbative
corrections \cite{Luis}, but not on complete quantizations of
gravitational models. It is hence far from clear whether the
conclusions about a minimum uncertainty can be maintained in a
nonperturbative quantization. With this motivation in mind, we
will study a very specific kind of models whose quantization can
be achieved both in a low-energy or perturbative scheme and by
taking gravity into full account.

We consider a dynamical system that, around a certain background
or in a certain approximation, can be described by a
time-independent Hamiltonian $H_0$ with associated time parameter
$T$. We assume that this system admits a straightforward
quantization. The quantum evolution of any explicitly
time-independent observable $A$ is dictated by the Heisenberg
equation $i\hbar\partial_T A=[A,H_0]$. For simplicity, we also
suppose that the spectrum of $H_0$ is positive and unbounded, with
a nondegenerate fundamental state. At this stage, we let general
relativity come into play in a fully nonperturbative way. Our
basic hypothesis, inspired by our introductory comments, is that
the main effect of plugging in gravity is changing the
normalization of the (asymptotic) time translations by an
energy-dependent factor. Since the physical time $t$ must be
normalized to the unity, we arrive at a relation of the form
$t=T\, V(H_0 E_P^{-1})$, where $V$ is a function on $I\!\!\!\,R^+$
and $E_P$ is a constant energy that can be viewed as a sort of
Planck energy for the system.

Since normalization factors are always positive, the function $V$
has to be greater than zero. In addition, to recover $T$ as the
time coordinate in the low-energy or perturbative limit in which
$H_0E_P^{-1}$ vanishes, we must have $V(0)=1$. Finally, we
introduce the assumptions that $V$ be monotonic and sufficiently
smooth to avoid technical complications.

Remarkably, one can construct a consistent quantization of the
system in general relativity starting from the quantization that
describes the evolution in the time $T$ with Hamiltonian $H_0$
\cite{BMV}. The Hilbert spaces of quantum states on the initial
$t=0$ and $T=0$ sections can be identified. Besides, one can check
that the evolution in the physical time $t$ is generated (at least
classically) by the Hamiltonian $H=E_P F(H_0E_P^{-1})$, where the
function $F$ is a primitive of $1/V$. As an operator, $H$ can be
defined from $H_0$ by means of the spectral theorem. The
explicitly time-independent observables satisfy now the equation
$i\hbar\partial_t A=[A,H]$. We choose $F(0)=0$, so that the ground
state energy vanishes also in the presence of gravity. Since
$F^{\prime}(x)=1/V(x)>0$ (because $V$ is positive and smooth), the
spectrum of $H$ is hence nonnegative. Apart from a factor $E_P$,
this spectrum coincides with the image under $F$ of that of
$H_0E_P^{-1}$.

From now on, we will refer to the quantizations with Hamiltonian
$H_0$ and $H$, respectively, as the perturbative and
nonperturbative quantizations. As a motivation for this
terminology, note that, since $H=E_P F(H_0E_P^{-1})$ and $F(0)=0$,
one can think of the perturbative approach as the analysis in the
limit $E_P^{-1}\rightarrow 0$, in agreement with our previous
comments. It can be seen that this analysis reproduces as well the
low-energy behavior $H_0\approx 0$.

One can doubt that a model of this type may represent a realistic
situation in general relativity. However, there is at least one
known example: the Einstein-Rosen (ER) waves. These cylindrical
gravitational waves are classically equivalent to a massless,
axisymmetric scalar field coupled to gravity in three dimensions
\cite{AP}. In linearized gravity, the corresponding
three-dimensional reduction of the metric (in a suitable gauge) is
purely Minkowskian, and the dynamics in this Minkowskian time $T$
is generated by the Hamiltonian $H_0$ of the free, massless scalar
field \cite{BMV}. Moreover, the time translations $\partial_T$ are
asymptotically unit even from the perspective of the
four-dimensional metric. In cylindrical general relativity without
any linearization, on the other hand, the metric in three
dimensions is not Minkowskian anymore and the physical time $t$,
properly normalized at infinity (both from the three and
four-dimensional viewpoints), differs from that of the Minkowski
background by a factor that depends on the energy of the free
field, $H_0$ \cite{BMV}.

Explicitly, $t=T e^{4G_3H_0}$ for ER waves, where the inverse
energy $G_3$ denotes the gravitational constant per unit length in
the direction of the axis or, equivalently, the effective Newton
constant in three-dimensions \cite{BMV,AP}. With our notation, we
then have $V(x)=e^{4x}$ and $E_P=1/G_3$. The primitive of $1/V$ is
$F(x)=(1-e^{-4x})/4$ and the Hamiltonian $H$ in the nonlinear
theory is $4G_3H=1-e^{-4G_3H_0}$. Thus, the physical energy ranges
from zero to $1/(4G_3)$.

Let us study the uncertainty in the time $t$ in our family of
models. The main observation, already pointed out in the analysis
of ER waves \cite{AP}, is that the physical time $t$ plays the
role of evolution parameter in the nonperturbative quantization,
whereas this role corresponds to the time $T$ in the perturbative
case \cite{BMV}. In this latter quantization, the physical time
$t=T V(H_0E_P^{-1})$ is represented by a one-parameter family of
operators. It seems natural to define $V(H_0E_P^{-1})$ in terms of
the Hamiltonian $H_0$ using the spectral theorem. The operator
obtained in this way is positive, because so is the function $V$.

The uncertainty in the nonperturbative quantization is
straightforward to analyze. Since the physical time is a dynamical
parameter, the fourth Heisenberg relation applies, i.e., $\Delta
t\Delta H\geq \hbar/2$. In the light of this relation, we arrive
at an unexpected conclusion. Namely, in the description of a fully
nonperturbative observer, the existence of a minimum time
uncertainty depends only on whether the rms deviation $\Delta H$
of the physical energy is or not bounded from above. Recalling
that the spectrum of $H_0$ is unbounded and the definition of $H$,
one can check that the largest that $\Delta H$ may become is $E_P
F_{\infty}$. Here, $F_{\infty}$ is the limit of $F(x)$ when $x$
tends to infinity. Therefore, a resolution limit exists in the
nonperturbative model if and only if the range of $F$ is bounded.
This happens to be the case for ER waves, where $F_{\infty}=1/4$,
and so one has $\Delta t\geq 2\hbar E_P^{-1}=2\hbar G_3$.
Nevertheless, in more general situations, nothing seems to prevent
the range of $F$ to be the whole positive semiaxis. The
uncertainty $\Delta t$ might then be decreased to zero by choosing
a state with totally uncertain energy $H$.

We now turn to the perturbative approach. Given a quantum state,
we can always measure on it the probability distribution of the
perturbative energy $H_0$, which is stationary because the system
is conservative. Via the spectral theorem, this distribution
determines that of the operator $V(H_0 E_P^{-1})$. We denote by
$\Delta V$ the corresponding rms deviation. In order to evaluate
the operator $t$, we still need to fix the value of the parameter
$T$. As we have explained above, we can detect the passage of the
time $T$ in the perturbative framework by examining the evolution
of probability distributions of observables in our quantum state.
This leads to a statistical measurement of the value of $T$, with
a distribution $\rho(T)$ whose uncertainty must be at least of the
order of $\Delta T\geq \hbar/(2\Delta H_0)$, according to
Heisenberg relation. Note that, in order to capture the intrinsic
uncertainties of the system, we choose to evaluate $T$ employing
indeed (different copies of) our state vector. Since the described
measurements of the perturbative energy $H_0$ and $T$ are
independent, our measurement procedure assigns to the physical
time $t$ a probability distribution which is the product of those
found for $T$ and $V$.

Remembering the stationarity of the energy, a straightforward
calculation then shows
\begin{eqnarray}
[\Delta t]^2&=& \int dT \;\rho(T) \;\langle\; T^2V^2-T_0^2 \langle
V\rangle^2\;\rangle\nonumber\\ \label{unc}&=&T_0^2[\Delta V]^2+
\langle V\rangle^2[\Delta T]^2+[\Delta T\Delta V]^2.\end{eqnarray}
Here, $T_0$ is the mean value of $T$ obtained with the
distribution $\rho(T)$, and $\langle\;\rangle$ denotes expectation
value (which can be computed employing the spectral resolution of
the identity and the probability distribution of $H_0$).

The above formula implies that the uncertainty in the physical
time cannot vanish in the perturbative quantization. To prove this
assertion notice that, in order that $\Delta t$ vanishes, the
three factors that appear in Eq. (\ref{unc}) must be zero. But, as
soon as $T_0\neq 0$, this can only occur if both $\Delta T$ and
$\Delta V$ vanish, because $V$ is a positive operator. On the
other hand, the spectrum of this operator is, by construction, the
image of the spectrum of $H_0E_P^{-1}$ under the function $V(x)$.
With the assumption that this function be monotonic, the vanishing
of $\Delta V$ guarantees that the analyzed state is an eigenvector
of $H_0$. However, owing to the fourth Heisenberg relation,
$\Delta T$ may vanish only if the quantum state has an infinite
uncertainty in the perturbative energy $H_0$. We thus arrive at a
contradiction. Hence, the uncertainty in the physical time, as
determined by an observer in the perturbative theory, must be
strictly positive except perhaps at the initial time of the
measurements.

It is instructive to analyze the consequences of Eq. (\ref{unc})
when one keeps only the first perturbative correction to the
prediction of ordinary quantum mechanics. Expanding
$V(H_0E_P^{-1})$ in powers of $E_P^{-1}$ and using $V(0)=1$,
\begin{equation}
V(H_0E_P^{-1})=1+V^{\prime}(0)H_0E_P^{-1} +O(
H_0E_P^{-1})^2.\end{equation}We then obtain $\langle V\rangle=1$
and $\Delta V=\Delta H_0\,E_P^{-1}|V^{\prime}(0)|$ at leading
order. Substituting this in Eq. (\ref{unc}),
\begin{equation}\label{uncp}
[\Delta t]^2=\Delta T^2+[\Delta H_0]^2 E_P^{-2}\;|V^{\prime}(0)|^2
\left(T_0^2 + \Delta T^2\right).\end{equation} Remembering that
$\Delta T\Delta H_0\geq \hbar/2$ and following a line of reasoning
similar to that employed to calculate the minimum uncertainty for
the SW clock \cite{AC,SW}, one concludes
\begin{equation}\label{tmin}
[\Delta t]^2\geq \frac{1}{4}
|V^{\prime}(0)|^2t_P^2+|V^{\prime}(0)|t_P T_0,\end{equation} where
$t_P=\hbar E_P^{-1}$ can be understood as the Planck time. It is
worth pointing out that the deduction of this equation is in fact
formally independent of the supposition about the monotonicity of
the function $V$.

Formula (\ref{tmin}) has a striking resemblance with the kind of
effective equation proposed in Ref. \cite{AC2} to describe the
limitation on the measurability of distances. The first term on
the right-hand side gives a constant uncertainty of the order of
the Planck time $t_P$, and can be interpreted as a quantum
uncertainty of pure gravitational origin, independent of the
details of the state employed in the measurement process
\cite{AC2}. The second contribution is an uncertainty of the order
of $\sqrt{t_P T_0}$, which has the same time dependence that is
found in SW devices or in random walk models of Planckian
fluctuations \cite{AC,SW}. It can be regarded as originated by the
quantum uncertainties that exist on the state used for the time
measurements.

We have thus shown that, for the type of models under study, the
fact that the physical time is represented as a one-parameter
family of operators in the perturbative theory, together with the
procedure by which these operators are measured, implies a
nonvanishing minimum time uncertainty, lending in this sense
confirmation to the perturbatively inspired analyses found in the
literature \cite{Luis,AC}. On the other hand, in a purely
nonperturbative quantization, the physical time can be assigned
the role of a dynamical parameter whose uncertainty is restricted
only by the standard Heisenberg relation. The time resolution can
then be improved without limit if the physical Hamiltonian is
unbounded.

Regarding the consequences on gravitational wave detectors and
stellar interferometry \cite{ACW,LH}, our main remark is that the
spacetime structure needs not be foamy in our models. Actually,
the uncertainty in the physical time (\ref{tmin}) emerges in the
perturbative quantization just from two independent processes: the
evaluation of $T$ and the measurements of the perturbative energy.
In interferometric experiments like those considered here,
moreover, an observer in the perturbative theory would register
the superposition of two simultaneous signals at the same instant
$T$ (which does not even need to be evaluated). Therefore, in our
particular class of models, this observer ought not to experiment
the kind of phenomena described in Refs.
\cite{AC,ACW,NgVD,LH,RTG}. Although a more detailed analysis and
specification of the system is required for definite predictions,
we hope that our discussion contributes to emphasize the relevance
of the measurement procedure and the hypothesis about the foamy
behavior of the spacetime (which was explicitly assumed in the
above references).

Let us conclude with some general comments. The first one refers
to the genuinely nonperturbative results of loop quantum gravity
about geometric operators, e.g., the area. The spectrum of these
operators is discrete \cite{area}, leading to a noncontinuous
spacetime picture at small scales. However, these results do not
necessarily imply a minimum spacetime resolution. In fact,
consecutive area eigenvalues are separated by a (square) distance
that vanishes as one approaches the sector of infinite large
areas, where an infinite resolution may be reached.

The other comment concerns the feasibility of our models. In Eq.
(\ref{uncp}) for the time uncertainty, the last term (proportional
to $\Delta H_0^2\Delta T^2$) and the first one ($\Delta T^2$) may
be interpreted as contributions with a purely quantum
gravitational origin and a standard quantum mechanical origin,
respectively \cite{AC2}. The remaining term should then provide
the leading gravitational correction emerging from the
uncertainties on the state of the system. One would hence expect
that the factor $(E_P^{-1}|V^{\prime}(0)|T_0)^2$ multiplying
$\Delta H_0^2$ in this term were independent of the Planck
constant $\hbar$, since it must originate from general relativity.
As a result, the associated energy $E_P$ should be independent of
$\hbar$. This is what actually occurs for ER waves, where
$E_P=1/G_3$. But in more general systems that do not admit
reduction to three dimensions, one expects $E_P$ to be given by
the quantum gravitational Planck scale $\sqrt{\hbar/G}$, where $G$
is (the four-dimensional) Newton constant. In this case, one might
argue that the fully nonlinear gravitational behavior will not
lead to the simple kind of effects assumed in our models. Even so,
the system may possess a scale that replaces $\sqrt{\hbar/G}$ in
our discussion and does not vanish when $\hbar=0$. An interesting
possibility is the presence of a cosmological constant $\Lambda$.
It might then happen that the role of $E_P$ could be assigned to
$1/\sqrt{|\Lambda| G^2}$ under certain circumstances.

The authors are grateful to A. Ashtekar, M. Varadarajan, and L.
Garay for enlightening conversations. This work was supported by
the Spanish MCYT projects BFM2002-04031-C02 and BFM2001-0213, and
E.J.S.V. by a Spanish MEC grant.

\end{document}